\documentclass[]{aa}

\usepackage{txfonts}

\usepackage{graphicx}

\begin{document}

\def\P{\bar{\Phi}}

\def\st{\sigma_{\rm T}}

\def\vk{v_{\rm K}}

\def\sles{\lower2pt\hbox{$\buildrel {\scriptstyle <}
   \over {\scriptstyle\sim}$}}

\def\sgreat{\lower2pt\hbox{$\buildrel {\scriptstyle >}
   \over {\scriptstyle\sim}$}}

\title{Could Dark Matter Interactions be an Alternative to Dark Energy ?}

\author{Spyros Basilakos$^1$ and Manolis Plionis$^2$}
\institute{$^1$ Research Center for Astronomy, Academy of Athens, 
GR-11527 Athens, Greece, \\
$^2$Institute of Astronomy \& Astrophysics, 
National Observatory of Athens, Thessio 11810, Athens, Greece \&
Instituto Nacional de Astrof\'isica, \'Optica y Electr\'onica, 72000 Puebla, Mexico
\\
\email{svasil@academyofathens.gr; mplionis@astro.noa.gr}}

\titlerunning{Dark matter interactions and dark energy}

\date{Received / Accepted }

\abstract
{
We study the global dynamics of the universe within the framework of the 
Interacting Dark Matter (IDM) scenario.
Assuming that the dark matter obeys the collisional Boltzmann
equation, we can derive analytical 
solutions of the global density evolution, which can accommodate an accelerated
expansion, equivalent to either the {\em quintessence} or the 
standard $\Lambda$ models, with the present time located after the inflection point.
This is possible if there is a
disequilibrium between the DM particle creation and annihilation processes
with the former process dominating, which creates an effective source
term with negative pressure. Comparing the predicted Hubble expansion
of one of the IDM models (the simplest) with observational data we find that the effective
annihilation term is quite small, as suggested by a variety of other
recent experiments.

}

\maketitle

\section{Introduction}
Over the past decade the analysis of high quality cosmological 
data (supernovae type Ia, CMB, galaxy clustering, etc.) have suggested that we live in a flat and 
accelerating universe, which contains cold dark matter to explain clustering and an extra 
component with negative pressure, the vacuum energy (or in a more general setting
the {\em dark energy}), to explain the observed accelerated cosmic expansion
(Spergel et al. 2007; Davis et al. 2007; 
Kowalski et al. 2008; Komatsu et al. 2009 and references therein).
Due to the absence of a physically well-motivated fundamental theory, there have 
been many theoretical speculations regarding the nature of the above exotic dark energy (DE)
among which a cosmological constant, scalar or vector fields
(see Weinberg 1989; Wetterich 1995; Caldwell, Dave \& Steinhardt 1998; Brax \& Martin 1999; 
Peebles \& Ratra 2003; Perivolaropoulos 2003;
Brookfield et al. 2006; 
Boehmer \& Harko 2007 and references therein).

Most of the recent papers in this kind of studies are based on the assumption that
the DE evolves independently of the dark matter (DM). The unknown nature
of both DM and DE implies that we can not preclude future surprises 
regarding the interactions in the dark sector. 
This is very important because interactions between the DM and 
{\em quintessence} could provide possible solutions 
to the cosmological coincidence problem
(Grande, Pelinson \& Sol\'a 2009). 
Recently, several papers have been published in this area 
(eg., Amendola et al. 2003; Cai \& Wang 2005; Binder \& Kremer 2006;
Campo et al. 2006; Wang, Lin \& Abdalla 2006; Das, Corasaniti, \& Khoury 2006;
Olivares, Atrio-Barandela \& Pavon; He \& Wang 2008, and references therein)
proposing that the DE and DM could be coupled, assuming also that 
there is only one type of non-interacting DM. 

However, there are other possibilities: (a) It is plausible that the dark
matter is self-interacting (IDM)
[Spergel \& Steinhardt 2000],
a possibility that has been proposed to solve discrepancies between 
theoretical predictions and astrophysical observations, 
among which the less cuspy halo profiles, predicted by the IDM model,
allowing for the observed gamma-ray and microwave emission 
from the center of our galaxy (Flores \& Primack 1994; Moore et al. 1999; 
 Hooper, Finkbeiner \& Dobler 2007; Regis \& Ullio 2008 and references
 therein) and the discrepancy between the predicted optical depth,
 $\tau$, from the Gunn-Peterson test in the spectra of high-z QSOs and
 the WMAP-based value (eg., Mapelli, Ferrara \& Pierpaoli 2006; 
Belikov \& Hooper 2009; Cirelli, Iocco \& Panci 2009
 and references therein). 
It has also been shown that some dark matter interactions 
could provide an accelerated expansion phase of the
Universe (Zimdahl et al. 2001; Balakin et al. 2003; Lima, Silva \& Santos 2008),
(b) The DM could potentially contain more than one particle species,
for example a mixture of cold and warm or hot dark matter
(Farrar \& Peebles 2004; Gubser \& Peebles 2004), with or without inter-component
interactions.

In this work we are not concerned with the viability of the different 
such possibilities, nor with the properties of interacting DM models.
The single aim of this work is to investigate whether there are repercussions of DM
self-interactions 
for the global dynamics of the universe and specifically whether
such models can yield 
an accelerated phase of the cosmic expansion, without the need of dark energy.
Note that we do not ``design'' the fluid interactions to
produce the desired accelerated cosmic evolution, as in some previous works
(eg., Balakin et al. 2003), but rather we investigate under which circumstances
the analytical solution space of the collisional Boltzmann equation, in
the expanding Universe, allows for a late accelerated phase of the
Universe.

\section{Collisional Boltzmann Equation in the Expanding Universe}
It is well established that the global dynamics of a homogeneous,
isotropic and flat Universe is given by the Friedmann
equation:
\begin{equation}
\left(\frac{\dot{\alpha}}{\alpha}\right)^2 = \frac{8 \pi G}{3} \rho \;,
\end{equation}
with $\rho$ the total energy-density of the cosmic fluid, containing
(in the matter dominated epoch) dark matter, baryons and any type of 
exotic energy. 
Differentiating the above, we derive the second Friedmann equation,
given by:
\begin{equation}\label{dda}
\frac{\ddot{\alpha}}{\alpha} = -\frac{4 \pi G}{3} \left(-2\rho -\frac{\dot \rho}{H} \right) \;\;.
\end{equation}

As we have mentioned in the introduction, the dark matter is usually
considered to contain only one type of particle that 
is stable and neutral. In this work we investigate, using
the Boltzmann formulation, the cosmological
potential of a scenario in which the dominant ``cosmic'' fluid 
does not contain dark energy, 
is not perfect and at the same time it is not in equilibrium
\footnote{Initially, the total energy density is $\rho=\rho_{\rm IDM}+\rho_{r}$. 
We consider that the self interacting dark matter does not interact significantly 
with the background radiation, and thus in the matter dominated epoch, radiation
is irrelevant to the global dynamics (due to the well-known dependence: 
$\rho_{r} \propto a^{-4}$). 
Therefore, taking the above considerations into account and assuming that there are no 
residual radiation products of the DM interactions (otherwise see Appendix A),
 we conclude that 
in the matter dominated era the total cosmic dark-matter density reduces 
to that of the IDM density ($\rho\simeq \rho_{\rm IDM}$) 
which obeys the collisional Boltzmann
equation (see eq. \ref{bol1}).}. 
Although our
approach is phenomenological, we will briefly review a variety of physically
motivated dark matter self-interaction models, which have appeared in
the literature.

The time evolution of the total density of the cosmic fluid is 
described by the collisional Boltzmann equation:
\begin{equation}
\frac{{\rm d}\rho}{{\rm d}t}+3H(t)\rho+\kappa \rho^{2}-\Psi=0 \;,
      \label{bol1} 
\end{equation}
where $H(t)\equiv \dot{\alpha}/\alpha$ is the Hubble function, 
$\Psi$ is the rate of creation of the DM particle pairs and
$\kappa (\ge 0)$ is given by:
\begin{equation}
\kappa=\frac{\langle \sigma u \rangle}{M_x} \;,
\end{equation}
where $\sigma$ is the cross-section for annihilation, $u$ is the mean particle velocity,
and $M_x$ is the mass of the DM particle.

Note that, in the context of a spatially flat FLRW cosmology, for 
an effective pressure term of:
\begin{equation}\label{PRES}
P=\frac{\kappa \rho^2-\Psi}{3 H}\;,
\end{equation}
the collisional Boltzmann equation reduces to the usual fluid
equation: $\dot{\rho}+ 3H (\rho + P)=0$. Inserting eqs.(\ref{bol1}) and
(\ref{PRES}) into eq.(\ref{dda}) we now obtain:
\begin{equation}\label{d1}
\frac{\ddot{\alpha}}{\alpha} = -\frac{4 \pi G}{3} \left(\rho +\frac{\kappa \rho^{2}-\Psi}{H} \right)= 
-\frac{4 \pi G}{3} \left(\rho +3P\right) \;\;.
\end{equation}
Obviously, a negative pressure (whichever its cause) can 
effectively act as a repulsive force possibly providing a cosmic acceleration.

In this paper we investigate the effects of DM self-interactions to
the global dynamics of the Universe and under which circumstances
they can produce a negative pressure and thus provide an alternative to
the usual dark energy.
It is well known that negative pressure implies tension rather
than compression, an impossibility for ideal gases, but not so for some physical
systems which depart from thermodynamic equilibrium 
(Landau \& Lifshitz 1985).

The particle annihilation regime has been described by
Weinberg (2008), using the collisional Boltzmann
formulation, in which the physical properties of the DM
interactions are related to massive particles (still being present)
which if they carry a conserved additive or multiplicative quantum
number,
would imply that some particles must be left over 
after all the antiparticles have annihilated (Weinberg calls them 
L-particles). 
The L-particles may annihilate to other particles, which during the period of annihilation
can be assumed to be in thermal and chemical equilibrium 
(see Weinberg 2008).
Such a DM self-interacting model has repercussions to the global
dynamics of the Universe (see our {\em Case 2} below).

The corresponding effects to the global dynamics of the particle
creation regime, providing an effective negative pressure, 
has also been investigated by a number of authors
(eg., Prigogine et al. 1989; Lima et al. 2008 and references therein).

Generally, in the framework of a Boltzmann formalism, a negative
pressure could indeed be the outcome 
of dark matter self-interactions, as suggested in 
Zimdahl et al. (2001) and Balakin et al. (2003),
if an ``antifrictional'' force is self-consistently exerted 
on the particles of the cosmic fluid.
This possible alternative to dark energy has the caveat of its unknown exact
nature, which however is also the case for all dark energy models.
Other sources of negative pressure have also been proposed, among which 
gravitational matter ``creation'' processes (Zeldovich 1970),
viewed through non-equilibrium thermodynamics (Prigogine et al. 1989)
or even the C-field of Hoyle \& Narlikar (1966). The effects of
the former
proposal (gravitational matter creation) on the global dynamics of the
Universe have been investigated, under the assumption that the
particles created are non-interacting (Lima et al. 2008).
The merit of all these alternative models is that they unify the dark sector 
(dark energy and dark matter), since just a single dark component (the
dark matter) needs to be introduced in the cosmic fluid. 

In what follows we present, in a unified manner, the outcome for the
global dynamics of the Universe of different type of dark matter
self-interactions, using the Boltzmann formulation in the matter dominated era.

\section{The Cosmic Density Evolution for different DM interactions}
We proceed to analytically solve eq. (\ref{bol1}). 
We change variables from $t$ to $\alpha$ and thus eq.(\ref{bol1}) 
can be written:
\begin{equation}
\frac{{\rm d}\rho}{{\rm d}\alpha}=f(\alpha)\rho^{2}+g(\alpha)\rho+R(\alpha) \;\;
      \label{bol2} 
\end{equation}
where
\begin{equation}
f(\alpha)=-\frac{\kappa}{\alpha H(\alpha)} \;\;\; g(\alpha)=-\frac{3}{\alpha} \;\;\;
R(\alpha)=\frac{\Psi(\alpha)}{\alpha H(\alpha)}\;.      
      \label{fun1} 
\end{equation}
Within this framework, based on eqs.(\ref{PRES}, \ref{bol2}) and (\ref{fun1}), we
can distinguish four possible DM self-interacting cases.

\underline{\it Case 1: $P=0$:}
If the DM is collisionless or the collisional annihilation and pair creation
processes are in equilibrium (ie., $\Psi \equiv \kappa \rho^{2}$), the corresponding solution of the 
above differential equation is $\rho \propto \alpha^{-3}$
(where $\alpha$ is the scale factor of the universe), and thus we
obtain, as we should, 
the dynamics of the Einstein de-Sitter model, with $H(t)=2/3t$.

\underline{\it Case 2: $P=\kappa \rho^{2}/3H$}:
If we assume that in the matter era the 
particle creation term is negligible,  
$\Psi=0$ [$R(\alpha)=0$], (the case discussed in Weinberg 2008),  
then the corresponding pressure becomes positive. 
It is clear that eq. (\ref{bol2}) becomes a Bernoulli equation, 
the general solution of which provides the evolution of the
global energy-density, which is that corresponding to the IDM ansanz:
\begin{equation}
      \label{bern} 
\rho(\alpha)= 
\frac{\alpha^{-3}}
{{\cal C}_2 - \int_{1}^{\alpha} x^{-3}f(x) dx }= 
\frac{\alpha^{-3}}
{{\cal C}_2+\kappa \int_{t_0}^{t} \alpha^{-3}(t) dt } \;\;.
\end{equation}
Prior to the present epoch we have that $\rho(\alpha) \propto
\alpha^{-3}$, while at late enough times ($\alpha \gg 1$) 
the above integral converges, which implies that the corresponding global density 
tends to evolve again as the usual dark matter (see Weinberg 2008), with
\begin{equation}
\rho(\alpha)\rightarrow
\frac{\alpha^{-3}}
{{\cal C}_2+\kappa \int_{t_{0}}^{\infty} \alpha^{-3}(t) dt } \propto \alpha^{-3}\;,
\end{equation}
where $t_{0}$ is the present age of the Universe.
The latter analysis, relevant to the usual weakly interacting 
massive particle case - Weinberg (2008), 
leads to the conclusion that the annihilation term 
has no effect resembling that of dark energy, but does affect 
the evolution of the self interacting DM component, with
the integral in the denominator rapidly converging to a constant (which
does depend on the annihilation cross-section).

\underline{\it Case 3: $P=(\kappa \rho^2-\Psi)/3 H$}:
For the case of a non-perfect DM fluid 
(ie., having up to the present time a disequilibrium
between the annihilation and particle pair creation processes) 
we can either have a positive or a negative effective
pressure term. 
Although the latter situation may or may not appear plausible, 
even the remote such possibility, 
ie., the case for which the DM particle creation term is larger 
than the annihilation term ($\kappa \rho^{2}-\Psi<0$),  is of
particular interest for its repercussions on the global dynamics of
the Universe (see for example 
Zimdahl et al. 2001; Balakin et al. 2003). 

It is interesting
to note that this case can be viewed as a generalization
of the gravitational matter creation
model of 
Prigogine et al. (1989) [see also Lima et al. 2008 and references therein]
in which annihilation processes are also included, although the
matter creation component dominates over annihilations.
In such a scenario, as well as in any interacting dark-matter model 
with a left-over residual radiation,
 a possible contribution from the radiation
products 
to the global dynamics is negligible, as we show in appendix A.

In general, for $\kappa \ne 0$ and $\Psi \ne 0$ it is not an easy task 
to solve analytically eq. (\ref{bol2}), which is a Riccati equation, 
due to the fact that it is a non-linear differential equation.
However, eq.(\ref{bol2}) could be 
fully solvable if (and only if) a particular solution is known. Indeed,
we find that for some special cases regarding the functional form of the
interactive term, such as $\Psi=\Psi(\alpha,H)$, we can derive 
analytical solutions.
We have identified two functional forms for which
we can solve the previous differential equation analytically, 
only one of which is of interest since it provides a
$\propto a^{-3}$ dependence of the scale factor (see appendix B). This is:
\begin{equation}
\Psi(\alpha)=\alpha H(\alpha)R(\alpha)={\cal C}_{1}(m+3)\alpha^{m}
H(\alpha)+\kappa {\cal C}_{1}^{2}\alpha^{2m} \;.
      \label{int1} 
\end{equation}  
Although, the above functional form was not
motivated by some physical theory, but rather 
phenomenologically by the fact that it provides analytical solutions
to the Boltzmann equation, its exact form can be justified {\em a posteriori}
within the framework of IDM (see appendix C).

The general solution of equation (\ref{bol2}) for the total
energy-density, using eq.(\ref{int1}), is:
\begin{equation}
\rho(\alpha)={\cal C}_{1}\alpha^{m}+\frac{\alpha^{-3}F(\alpha)}
{\left[{\cal C}_2-\int_{1}^{\alpha} x^{-3} f(x) F(x)dx\right]} \;,
      \label{sol1} 
\end{equation}  
where the kernel function $F(\alpha)$ has the form:
\begin{equation}
F(\alpha)={\rm exp}
\left[-2 \kappa {\cal C}_{1}\int_{1}^{\alpha} \frac{x^{m-1}}{H(x)}dx \right] \;\;.
      \label{ker2} 
\end{equation}  
Note that $\kappa {\cal C}_1$ has units of Gyr$^{-1}$, while 
$m$, ${\cal C}_{1}$ and ${\cal C}_2$ are the corresponding constants of the problem. 
Obviously, eq.(\ref{sol1}) can be seen as  
\begin{equation}
\rho(\alpha)=\rho_{c}(\alpha)+\rho^{'}(\alpha) \;,
      \label{gg33} 
\end{equation}
where $\rho_{c}=C_{1}\alpha^{m}$ is the density corresponding to the
residual ''matter creation'',
resulting from a possible 
disequilibrium between the particle creation and annihilation processes,
while $\rho^{'}$ can be viewed as the energy density of the self-interacting dark
matter particles which are dominated by the annihilation processes.
This can be easily understood if we set the constant ${\cal C}_{1}$ 
strictly equal to zero, implying that the creation term is negligible,
which reduces 
the current solution (eq.\ref{gg33}) to that of eq.(\ref{bern}).
Note that near the present epoch as well as at late
enough times ($\alpha \gg 1$), as also in {\em Case 2},  
the $\rho^{'}$ evolves as the usual dark matter (see also Weinberg 2008).
Finally, if both $\kappa$ and $\Psi$ tend to zero, the 
above cosmological model reduces to the usual Einstein-deSitter model
({\em Case 1}).

Note that, due to $\rho^{'}>0$, the constant ${\cal C}_{2}$ obeys the following restriction:
\begin{equation}
{\cal C}_{2}> G(\alpha)=\int_{1}^{\alpha} x^{-3} f(x) F(x)dx \ge 0 \;\;.
      \label{regg33} 
\end{equation}
Evaluating now eq.(\ref{sol1}) at the present time ($\alpha=1$, $F(\alpha)=1$), we obtain 
the present-time total cosmic density, which is: 
$\rho_0={\cal C}_{1}+1/{\cal C}_2\;$, with ${\cal C}_{1} \ge 0$ and ${\cal C}_{2}>0$. 

\underline{\it Case 4: $P=-\Psi/3H$}:
In this scenario we assume that the annihilation term is 
negligible [$\kappa=0$ and $f(\alpha)$=0] and the particle creation
term dominates. Such a situation is mathematically equivalent to the
gravitational DM particle creation process within 
the context of non-equilibrium thermodynamics  
Prigogine et al. (1989),
the important cosmological repercussions of which have been studied in
Lima et al. (2008 and references therein).
Using our nomenclature and $\kappa=0$, eq.(\ref{bol2}) 
becomes a first order linear differential equation, a general solution of which is:
\begin{equation}
\rho(\alpha)=\alpha^{-3}\left[\int_{1}^{\alpha} x^{3} R(x) dx+
{\cal C}_{2}\right] \;\;.
      \label{sol7} 
\end{equation}
The negative pressure can yield a late accelerated phase of the cosmic 
expansion (as in Lima et al. 2008), 
without the need of the required, in the ``classical'' cosmological models, 
dark energy.

Below, we investigate the conditions under which 
eqs. (\ref{sol1}) and (\ref{sol7}) could provide accelerating solutions, 
similar to the usual dark energy case.

\section{Case 3: $P=(\kappa \rho^2-\Psi)/3 H$}

\subsection{Conditions to have an inflection point and galaxy formation}
In order to have an inflection point at $\alpha=\alpha_{I}$ we must
have $\ddot{\alpha}_{I}=0$ (see eq. \ref{d1}).
The latter equality implies that $\rho+3P=0$ 
should contain a real root in the interval: $\alpha \in (0,1)$. 
Therefore, with the aid of eq.(\ref{sol1}),
(\ref{PRES}) and (\ref{int1}), we derive the following condition:
\begin{equation}\label{ZZ}
\frac{\alpha^{-3}F(H+2\kappa {\cal C}_{1}\alpha^{m})}{{\cal C}_{2}-G}+
\frac{\kappa \alpha^{-6}F^{2}}{({\cal C}_{2}-G)^{2}}-
(m+2){\cal C}_{1}\alpha^{m}H =0\;,
\end{equation}
from which we obtain that $m>-2$ (where ${\cal C}_1>0$, $\kappa
\ge 0$ and ${\cal C}_2-G>0$).
Evidently, if we parametrize the constant $m$ according to
$m=-3(1+w_{\rm IDM})$, we obtain the condition: $w_{\rm IDM}<-1/3$,
which means that the current cosmological model can be viewed as
a viable {\em quintessence} dark-energy look-alike, as far as the global dynamics is concerned.
Indeed, we remind the reader that the same restriction holds
for the usual dark energy model in which 
$P_{Q}=w \rho_{Q}$ ($w=const$).
Since the avenue by which the IDM model provides cosmic acceleration may appear
slightly involved, we present in appendix D its correspondence 
to the usual dark energy models.

Furthermore, in order to have growth of spatial density fluctuations, 
the effective DM part should be capable of clustering and providing the formation of galaxies, while
the effective dark energy term should be close to homogeneous.
Indeed in our case the effective term that acts as dark energy is homogeneous in the
same sense as in the classical quintessence, while the  $\kappa \rho^2$ term
slightly modifies the pure DM evolution.
In any case the interacting DM term after the inflection point tends
to an evolution $\propto a^{-3}$. 
During the galaxy formation epoch at high-$z$'s we
expect (due to the functional form of the DM term) that the slope of
the interacting DM term 
is not far from the classical DM evolution (we will explore
further these issues in a forthcoming paper).

\subsection{Relation to the Standard $\Lambda$ Cosmology}
As an example, we will show that for $m=0$ (or $w_{\rm IDM}=-1$) 
the global dynamics, provided by eq.(\ref{sol1}), is
equivalent to that of the traditional $\Lambda$ cosmology. To this end we use
${\rm d}t={\rm d}\alpha/ (\alpha H)$ and the basic kernel (eq. \ref{ker2}) 
becomes:
\begin{equation}
F(\alpha)={\rm exp}
\left[-2 \kappa {\cal C}_{1}\int_{1}^{\alpha} \frac{1}{xH(x)}dx \right]=
e^{-2\kappa {\cal C}_{1}(t-t_{0})}
      \label{ker4} 
\end{equation}  
where $t_{0}$ is the present age of the universe.
In addition, the integral in 
equation (\ref{sol1}, see also eq.\ref{regg33}) takes now the following form:
$G(\alpha)=-\kappa Z(t)$ and 
$Z(t)=\int_{t_{0}}^{t} \alpha^{-3}e^{-2 \kappa {\cal C}_1 (t-t_0)}$.
Note that at the present time we have $G(1)=0$.
Therefore, using the above formula, 
the global density evolution (eq. \ref{sol1}) can be written:
\begin{equation}
\rho(\alpha)={\cal C}_{1}+\alpha^{-3}\frac{e^{-2\kappa {\cal C}_{1}(t-t_{0})}}
{\left[{\cal C}_2-G(\alpha) \right]} \;\;.
      \label{sol4a} 
\end{equation}  
As expected, at early enough times ($t \rightarrow 0$) the overall density 
scales according to:
$\rho(\alpha) \propto a^{-3}$,  while close to the 
present epoch 
the density evolves according to:
\begin{equation}\label{eq:lam}
\rho(\alpha)\simeq {\cal C}_{1}+\frac{\alpha^{-3}}{{\cal C}_2} \;\;,
\end{equation}
which is approximately the corresponding evolution in the 
$\Lambda$ cosmology, in which the term ${\cal C}_1$ acts as the constant-vacuum term ($\rho_{\Lambda}$) 
and the $1/{\cal C}_2$ term acts like matter ($\rho_m$).

Note that the effective pressure term (eq. \ref{PRES}),
for $\kappa\rightarrow 0$, becomes:
$\Psi \sim 3 {\cal C}_1 H$, which implies that: $P\sim -\Psi/3 H =  -{\cal C}_1$.
Therefore, this case relates to the traditional 
$\Lambda$ cosmology, since ${\cal C}_1$ corresponds to 
$\rho_{\Lambda}$ (see eq. \ref{eq:lam}).
We now investigate in detail the dynamics of the $m=0$ 
model.

\begin{figure}
\includegraphics[angle=0,scale=0.465]{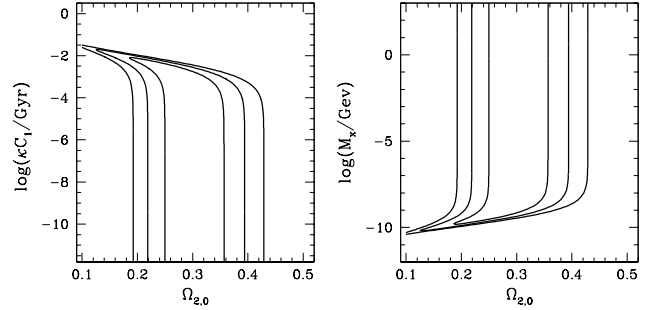}
\caption{{\em Left Panel}: The $\Omega_{2,0}-\kappa {\cal C}_1$ solution
  space provided by fitting our model to the early-type galaxy Hubble
  relation of Simon et al (2005). {\em Right Panel:} The corresponding
  $\Omega_{2,0}-M_x$ solution space.}
\end{figure}

From eq.(\ref{sol4a}), using the usual unit-less $\Omega$-like 
parameterization, we have after some algebra that:
\begin{equation}\label{hubb2}
\left(\frac{H}{H_0}\right)^2 = \Omega_{1,0} + \frac{\Omega_{1,0} \Omega_{2,0} \alpha^{-3} 
e^{-2 \kappa {\cal C}_1 (t-t_0)} }{\Omega_{1,0} + \kappa {\cal C}_1 \Omega_{2,0} Z(t)} \;,
\end{equation}
with $\Omega_{1,0}= 8 \pi G {\cal C}_1/3H_0^2$ and $\Omega_{2,0}=8 \pi G / 3H^2_0 {\cal C}_2$, 
which in the usual $\Lambda$ cosmology they correspond to $\Omega_{\Lambda}$ and 
$\Omega_{\rm m}$, respectively. 

We can now attempt to compare
the Hubble function of eq.(\ref{hubb2}) to that corresponding 
to the usual $\Lambda$ model.
To this end, we use a $\chi^2$ minimization between the different
models (our IDM eq.\ref{hubb2} or the traditional $\Lambda$CDM model) 
and the Hubble relation derived 
directly from early type galaxies at high redshifts 
(Simon,Verde, \& Jimenez 2005). For the case of our IDM model we
simultaneously fit the two free
parameters of the model, ie., $\Omega_{2,0}$ and $\kappa {\cal C}_1$
for a flat background ($\Omega_{1,0}=1-\Omega_{2,0}$) with
$H_{0}=72$ km/sec/Mpc and $t_0=H_0^{-1}\simeq 13.6$ Gyrs (roughly the age of the universe of
the corresponding $\Lambda$ cosmology).
This procedure yields, as the best fitted parameters, the following:
$\Omega_{2,0}=0.3^{+0.05}_{-0.08}$ and $\log(\kappa {\cal C}_1)\simeq -9.3$
(with stringent upper limit $\simeq -3$, but unconstrained towards lower values)
with $\chi^2/{\rm d.f.}= 1.29$ (see left panel of Fig. 1).
Using eq.(4) we can now relate the range of values of $\kappa {\cal C}_1$
with the mass of the DM particle from which we obtain that:
\begin{equation}
M_x=\frac{1.205 \times 10^{-12}}{\kappa {\cal C}_1} \frac{\langle \sigma u
  \rangle}{10^{-22}} \; {\rm GeV} \;,
\end{equation}
(see also right panel of Fig. 1) 
and since $\kappa {\cal C}_1$ is unbound towards small values, it is
consistent with currently 
accepted lower bounds of $M_x (\sim 10 {\rm GeV})$
(eg., Cirelli et al. 2009 and references therein). 
The corresponding Hubble relation (Fig. 2), provided by the best fitted
free parameters, is indistinguishable from that of the traditional
$\Lambda$CDM model, due to the very small value of $\kappa {\cal C}_1
\simeq 10^{-9.3}$.
For completion we also show, as the dashed line, the IDM solution 
provided by $M_x \sim 1$eV ($\kappa {\cal C}_1 \simeq 10^{-3}$), 
which is the stringent lower bound found
by our analysis. In this case the predicted Hubble expansion
deviates significantly from the traditional $\Lambda$ model 
at small $\alpha$ values indicating that it would probably create
significant alterations of the standard BBN 
(eg. Iocco et al. 2009 and references therein).

\begin{figure}
\includegraphics[angle=0,scale=0.44]{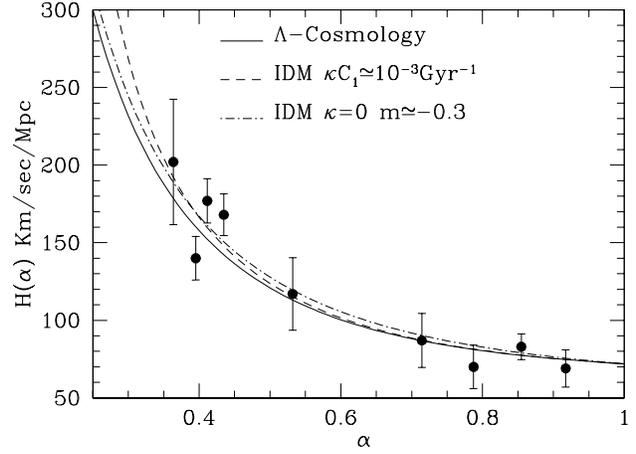}
\caption{Comparison of the Hubble function provided by the traditional
  $\Lambda$CDM model, which coincides with our $m=0$ model 
(for the best fit of the two free parameters - see text). 
The dashed line corresponds to our $m=0$ IDM model for the highest
$\kappa {\cal C}_1$ bound,
provided by our fitting procedure ($\sim 10^{-3}$).
The dot-dashed line corresponds to our $\kappa=0$ IDM model ({\em
  Case 4}) for the best fitted parameters ($m\simeq -0.3$ and
$\Omega_{2,0}\simeq 0.28$).
Finally, the points correspond to the observational data (Simon et al. 2005).}
\end{figure}

Although the present analysis does not provide any important constraints on $M_x$
(within our model), we plan to use in the future a large number of cosmologically relevant
data to attempt to place stronger $M_x$ constraints, also for the
general case of section 4.1.

\section{Case 4: $P=-\Psi/3H$}
In this section we will prove that for $\kappa=0$ (negligible annihilation),
the global dynamics resembles that of the traditional quintessence
cosmology ($w=$constant). Indeed, using again the phenomenologically 
selected form of $\Psi$, provided by eq.(\ref{int1}), we have 
$R(\alpha)={\cal C}_{1}(m+3)\alpha^{m-1}$. It is then straightforward
to obtain the density evolution from eq.(\ref{sol7}), as:
\begin{equation}
\rho(\alpha)={\cal D}\alpha^{-3}+{\cal C}_{1} \alpha^{m} \;,
      \label{sol77} 
\end{equation} 
where ${\cal D}={\cal C}_{2}-{\cal C}_{1}$.
The conditions under 
which the current model acts as a quintessence cosmology, are:
${\cal D}>0$, ${\cal C}_{1}>0$ and 
$w_{\rm IDM}=-1-m/3$, which implies that in order to have an inflection point,
the following should be satisfied: $w_{\rm IDM}<-1/3$ or $m>-2$ (see appendix D).
Notice, that the Hubble flow is now given by:
\begin{equation}
\left(\frac{H}{H_0}\right)^2 = \Omega_{\rm 2,0} \alpha^{-3}+ \Omega_{1,0} \alpha^{m} 
\end{equation}
with $\Omega_{\rm 2,0}= 8 \pi G {\cal D}/3H_0^2$ and $\Omega_{1,0}=8
\pi G {\cal C}_1/ 3H^2_0$.
Finally, by minimizing the corresponding $\chi^{2}$ (as in section 4.2), 
we find that the best fit values are $\Omega_{\rm 2, 0}\simeq 0.28$
and $m\simeq -0.30$ ($w_{\rm IDM}\simeq -0.90$) with $\chi^2/{\rm d.f.}= 1.29$. 
The corresponding Hubble flow curve is shown in Fig. 2 as
the dot-dashed line. Note that
this solution is mathematically equivalent to that of the gravitational matter
creation model of Lima et al. (2008).

\section{Conclusions}
In this work we investigate the evolution of the global 
density of the universe in the framework of an interacting DM scenario by solving
analytically the collisional Boltzmann equation in an expanding Universe. 
A disequilibrium between the DM particle creation and
annihilation processes, whichever its cause and in favor of the particle creation term,
can create an effective source
term with negative pressure which, acting like dark energy, provides
an accelerated expansion phase of the universe. There are also 
solutions for which the present time is located after the inflection point.
Finally, comparing the observed Hubble function of a few high-redshift elliptical
galaxies with that predicted by our simplest IDM model ($m=0$), we find that the effective
annihilation term is quite small.
In a forthcoming paper we will use a multitude of cosmologically
relevant observations to jointly fit the predicted, by our generic IDM
model, Hubble relation and thus possibly provide more stringent
constraints to the free parameters of the models. We plan also to 
derive the perturbation growth factor in order to
study structure formation within the IDM model.

\acknowledgements{}
We thank P.J.E. Peebles for critically reading our paper 
and for useful comments. 
Also, we would like to thank the anonymous referee
for his/her useful comments and suggestions.

\section*{Appendix A:The effect of the decay products}
Here we attempt to investigate in the matter-dominated era, 
whether the possible radiation 
products due to dark matter interactions can affect the global dynamics.
A general coupling can be viewed by the continuity equations of 
interacting dark matter $\rho_{\rm IDM}$ and 
residual radiation $\delta \rho_{r}$,

\begin{equation}
\frac{{\rm d}\rho_{\rm IDM}}{{\rm d}t}+3H(t)\rho_{\rm IDM}+\kappa \rho^{2}_{\rm IDM}-\Psi=Q \;,
      \label{abol1} 
\end{equation}

\begin{equation}
\frac{{\rm d}\delta \rho_{r}}{{\rm d}t}+4H(t)\delta \rho_{r}=-Q 
      \label{arad1} 
\end{equation}
where $Q$ is the rate of energy density transfer. If $Q<0$ then 
the IDM fluid transfers to residual radiation. 
As an example we can use a generic model
with $Q=-\epsilon \delta \rho_{r}$, where 
$\epsilon>0$. Thus, eq.(\ref{arad1}) has an exact solution 
\begin{equation}
\delta \rho_{r}=\delta \rho_{r0} \alpha^{-4} e^{\epsilon (t-t_0)}\;
\end{equation}
with $t_0$ the present age of the Universe.
This shows that the contribution of the residual radiation 
to the global dynamics is negligible in the past, 
since there is not only the usual $\propto a^{-4}$
dependence of the background radiation but also a further exponential drop, 
and thus $Q\simeq 0$
Therefore we conclude that we can approximate the total energy-density 
with that of the interacting dark-matter density
($\rho \simeq \rho_{\rm IDM}$).
Note, that $1/\epsilon$ can be viewed as the mean life time of the residual radiation
particles.

\section*{Appendix B: Solutions of the Riccati equation}
With the aid of the differential equation theory we present
solutions that are relevant to our eq.(\ref{bol2}). In general a Riccati 
differential equation is given by 
\begin{equation}
y^{'}=f(x)y^{2}+g(x)y+R(x)
\end{equation}
and it is fully solvable only when a particular solution is known.
Below we present two cases in which analytical solutions are possible.

\begin{itemize}
\item {\it Case 1:} For the case where:

\begin{equation}
R(x)={\cal C}_{1}mx^{m-1}-{\cal C}_{1}^{2}x^{2m}f(x)-{\cal C}
_{1}x^{m}g(x)
\end{equation}
the particular solution is $x^{m}$ and thus the corresponding
general solution can be written as:
\begin{equation}
y(x)={\cal C}_{1}x^{m}+\Phi(x)\left[{\cal C}_{2}-\int_{1}^{x} f(u)\Phi(u)du \right]^{-1}
\end{equation}
where
\begin{equation}
\Phi(x)={\rm exp}\left[\int_{1}^{x} \left(2 {\cal C}_{1}u^{m}f(u)+g(u)\right)du \right]
\end{equation}
and ${\cal C}_{1}, {\cal C}_{2}$ are the integration constants.
Using now eq.(\ref{fun1}) we get 
$\Psi(x)=xH(x)R(x)={\cal C}_{1}(m+3)x^{m} H(x)+\kappa {\cal C}_{1}^{2}x^{2m}$.

\item {\it Case 2:} For the case where:
\begin{equation}
R(x)=h^{'}(x) \;\;\;\; {\rm with}\;\;\;\;g(x)=-f(x)h(x)
\end{equation}
the particular solution is $h(x)$ [in our case we have $h(x)=-3\kappa^{-1}H(x)$].  
The general solution now becomes:
\begin{equation}
y(x)=h(x)+\Phi(x)\left[{\cal C}_{2}-\int_{1}^{x} f(u)\Phi(u) du \right]^{-1}
\end{equation}
where
\begin{equation}
\Phi(x)={\rm exp}\left[\int_{1}^{x} f(u)h(u) du\right] \;\;.
\end{equation}
In this framework, using eq. (\ref{fun1}) we finally get 
$\Psi(x)=xH(x)R(x)=-3\kappa^{-1}xH(x)H^{'}(x)$.
\end{itemize}
Note that the solution of {\it Case 1} 
is the only one providing a $\propto \alpha^{-3}$ dependence of the scale factor 
(see eqs. \ref{sol1}, \ref{sol4a} and \ref{eq:lam}).

\section*{Appendix C: Justification of functional form of  $\Psi$}
Suppose that we have a non-perfect cosmic fluid in a  
disequilibrium phase with energy density
$\rho$. Then from the collisional Boltzmann equation,
we have:
\begin{equation}
\Psi=\dot{\rho}+3H \rho+\kappa \rho^{2}=
\frac{{\rm d} \rho}{{\rm d}a} aH+3H \rho+\kappa \rho^{2}\;\;.
\end{equation}
Furthermore, we assume that for a convenient period of time
the cosmic fluid, in an expanding Universe, is slowly diluted according to 
$\rho \sim C_{1}\alpha^{m}$ ($m\le 0$). From a mathematical point of view, 
the latter assumption simply means 
that a solution of the form $\propto \alpha^{m}$ is a particular solution 
of the Boltzmann equation. Therefore, we have finally that:
\begin{equation}
\Psi \simeq C_{1}(m+3)a^{m}H+\kappa C^{2}_{1}a^{2m} \;\;.
      \label{bol7} 
\end{equation}

\section*{Appendix D: Correspondence between our model and the usual Dark Energy models}
We remind the reader that for homogeneous and 
isotropic flat cosmologies ($\Omega_{\rm m}+\Omega_{Q}=1$), driven by non
relativistic DM and a DE with a constant equation of state parameter ($w$), 
the density evolution of this cosmic fluid can be written as: 
\begin{equation}
\rho(\alpha)=\rho_{\rm m,0}\alpha^{-3}+
\rho_{Q,0} \alpha^{-3(1+w)} 
      \label{sol4} 
\end{equation} 
where $\rho_{\rm m,0}$ and $\rho_{Q,0}$ are 
the present-day DM and DE densities, respectively.

The necessary criteria to have cosmic acceleration and an inflection point 
in our past ($t_{i}<t_{0}$), are: (a) $P<0$ and (b) $\ddot{\alpha}=0$,
which leads to the conditions:
\begin{itemize}
\item {\bf Dark Energy models}: $P=P_{m}+P_{Q}=w \rho_{Q}<0$, $P_{m}=0$ with $w<-1/3$.
\item {\bf IDM models}: $P=\kappa \rho^2-\Psi <0$ and $m>-2$ (or
  $w_{\rm IDM}<-1/3$ see section 4).
\end{itemize}


\begin{thebibliography}{}
\bibitem[]{} Amendola L., Quercellini C., Tocchini-Valentini D., \& Pasqui A., 2003, ApJ, 583, L53 
\bibitem[]{} Balakin A.B., Pavon D., Schwarz D.J., \& Zimdahl W.,
  2003, N.J.Phys, 5, 85
\bibitem[]{} Belikov, A.V. \& Hooper, D., 2009, {\tt arXiv:0904.1210}
\bibitem[]{} Binder J.B., \& Kremer G.M., 2006, Gen. Rel. Grav. 38, 857  
\bibitem[]{}Boehmer C. G., \&, Harko T., 2007, Eur. Phys. J. C50, 423
\bibitem[]{}Brax P., \&, Martin J., 1999, Phys. Lett. 468, 40
\bibitem[]{}Brookfield A. W., van de Bruck C., Mota D. F., \& 
Tocchini-Valentini D., 2006, Phys. Rev. Lett., 96, 061301
\bibitem[]{}Caldwell R. R., Dave R., \&, Steinhardt P. J., 
1998, Phys. Rev. Lett., 80, 1582
\bibitem[]{}Cai R.G., \& Wang A., 2005, JCAP, 0503, 002
\bibitem[]{}Campo, R., Herrera, R., Olivares, G., \& Pavon, D., 2006,
  Phys. Rev. D., 74, 023501
\bibitem[]{} Cirelli, M., Iocco, F., Panci, P., 2009, {\tt arXiv:0907.0719}
\bibitem[]{}Das, S., Corasaniti, P.S., \& Khoury, J., 2006, Phys. Rev. D. 73, 083509
\bibitem[]{}Davis, T. M., et al., 2007, ApJ, 666, 716
\bibitem[]{}Farrar G.R., \& Peebles P.J.E. 2004, ApJ, 604 1  
\bibitem[]{}Flores R.A., \& Primack J.R., 1994, ApJ, 427, L1  
\bibitem[]{}Grande, J., Pelinson A., \&, Sol\'a, J., 2009, Phys. Rev D., 79, 3006
\bibitem[]{}Gubser S.S., \& Peebles P.J.E, 2004, Phys. Rev., D70, 123510 
\bibitem[]{}He, J.-H. \& Wang, B., 2008, JCAP, 06, 010
\bibitem[]{} Hooper D., Finkbeiner D.P., \& Dobler G., 2007, Phys. Rev. D., 76, 3012
\bibitem[]{}
Hoyle F., \&, Narlikar J.V., 1966, Proc, Roy, Soc. A, 290, 143 
\bibitem[]{}Komatsu, E., et al., 2009, ApJS, 180, 330
\bibitem[]{}Kowalski, M., et al., 2008, ApJ, 686, 749
\bibitem[]{}Landau L.D., \&, Lifshitz E. M., {\em Statistical Physics} (Pergamon ,Oxford, 1985)
\bibitem[]{}Lima J.A.S., Silva F.E., \& Santos R.C., 2008, 
Class. \& Quantum Gravity, 25, 205006
\bibitem[]{} Iocco, F., Mangano, G. Miele, G. Pisanti, O., Serpico,
  P.D., Phs.Rept., 2009, 472 
\bibitem[]{} Mapelli, M., Ferrara, A., Pierpaoli, E., 2006, MNRAS,
  369, 1719
\bibitem[]{} Moore B. et al. 1999, ApJ, 524, L19 
\bibitem[]{}Olivares, G., Atrio-Barandela, F., \& Pavon, D., 2008, Phys. Rev. D., 77, 063513
\bibitem[]{}Peebles, P. J. E.,\&  Ratra, B., Rev.Mod.Phys., 2003, 75, 559
\bibitem[]{}Perivolaropoulos L., 2003, Phys. Rev. D., 67, 1351
\bibitem[]{}Prigogine et al. 1989, Gen. Rel.Grav. 21, 767
\bibitem[]{}Regis M., \& Ullio P., 2008, {\tt astro-ph/0802.0234}
\bibitem[]{}Simon J., Verde L., \& Jimenez R., 2005, Phys. Rev. D., 71, 123001
\bibitem[]{}Spergel D.N., \& Steinhardt P.J. 2000, Phys. Rev. Lett., 84, 3760
\bibitem[]{}Spergel, D. N., et al., 2007, ApJS, 170, 377
\bibitem[]{}Wang, B., Lin, C.-Y., \& Abdalla, 2006, Phys. Lett. B., 637, 357
\bibitem[]{}Weinberg S., 1989, Rev. Mod. Phys., 61, 1
\bibitem[]{}Weinberg S., 2008, {\em Cosmology}, Oxford University Press
\bibitem[]{}Wetterich C., 1995, A\&A, 301, 321
\bibitem[]{}Zimdahl W., Schwarz D.J., Balakin A.B.,\& Pavon D., 2001, Phys. Rev.  D., 64, 3501
\bibitem[]{}Zeldovich Ya. B., 1970, JETP Lett. 12, 3007


\end{thebibliography}
\end{document}